\documentclass[twoside]{article}
\usepackage{fleqn,espcrc2}

\usepackage{graphicx}
\usepackage[figuresright]{rotating}

\newcommand{\AmS}{{\protect\the\textfont2
  A\kern-.1667em\lower.5ex\hbox{M}\kern-.125emS}}

\title{ Neutrino Masses and SO(10) SUSY GUTs }

\author{Stuart Raby\address{Department of Physics, The Ohio State University, 
174 W. 18 th Ave, Columbus, Ohio 43210, USA}%
\thanks{On leave at Universitat Bonn, Physikalisches Institut, Nussallee 12, 53115 Bonn, Germany.  The work presented in this talk was done in collaboration with T. Bla\v{z}ek and K. Tobe.   Partial support for this work came from an Alexander von Humboldt Fellowship and a DOE grant DOE/ER/01545.}}
       
\begin{document}

\begin{abstract}
In this talk I discuss the qualitative features of an SO(10) SUSY GUT with an SU(2)x U(1)$^n$ family symmetry.   I then describe the global fit of this theory to precision electroweak data [including charged fermion masses and mixing angles].  Finally, the predictions of the model for neutrino masses and mixing are discussed.   The most predictive version of the model naturally fits atmospheric neutrino data with maximal $\nu_\mu \longrightarrow \nu_\tau$ oscillations and solar neutrino data with small mixing angle MSW $\nu_e \longrightarrow \nu_{sterile}$ oscillations.
\vspace{1pc}
\end{abstract}

\maketitle

\section{ The Steps in Model Building }

\noindent
Before I discuss a particular model, let me first describe the steps one must take when constructing a candidate theory of nature.
\begin{itemize}
\item{\bf List the problems}

The initial step in model building requires one to analyze the experimental data and list the problems/questions left unresolved in the Standard Model.  For
example, we have the phenomenological/theoretical problems
\begin{enumerate}
\item Gauge hierarchy -- $M_Z/M_{Planck} \sim 10^{-17}$
\item Pattern of fermion masses and mixing
\item Absence of large flavor violation
\end{enumerate}

\item{\bf List the problem solving mechanisms}

The next step is to obtain a list of possible mechanisms for solving the above problems.  Each mechanism is capable of solving at least one (but preferably more than one) problem.   Note, many known mechanisms are mutually exclusive.

\item{\bf Develop a theoretical framework}

Finally, one develops a theoretical framework incorporating a maximal set of {\it mutually consistent} mechanisms.
\end{itemize}

Thus prior to presenting any specific model, let us consider the list of [{\bf mechanisms/problems}] we hope to incorporate into our model.
\begin{itemize}
\item  SUSY -  {\em a mechanism for} solving the gauge hierarchy problem.
\item GUTs -  {\em for} explaining charge quantization; family structure and reducing the number of fundamental parameters.
\item  Textures - {\em for} obtaining the pattern of fermion masses and mixing.
\item Family Symmetry - {\em for} explaining textures; the hierarchy of family masses and solving the flavor problem.
\end{itemize}

\section{ The Model }

The model is an
$SO_{10}$ SUSY GUT  with an $SU_2 \times U_1^n$ family symmetry\cite{1,3,4}.
The three families of quarks and leptons belong to the spinor representations
$ 16_a, \;\;\;  16_3$
where $ a = 1, 2 $ is an $SU_2$ family index.

Effective 3 x 3 Yukawa matrices are obtained at the GUT scale after spontaneously breaking the family symmetry and integrating out states
with mass  ($M, \; M_\chi$)  above $M_{G}$.  Schematically they have the form

$$\left(\begin{array}{ccc}  0 &  \epsilon^\prime  D  & 0 \\
- \epsilon^\prime  D  & \epsilon  C & \epsilon  B \\
 0 &  \epsilon  B  &   A \end{array} \right)$$

\noindent
where the small parameters   $$  \epsilon =   \frac{\langle \phi^2 \rangle \langle 45 \rangle }{ M M_\chi} \approx \frac{\langle S^{22} \rangle \langle 45 \rangle }{M M_\chi} $$
$$  \epsilon^\prime =    \frac{\langle A^{12} \rangle}{M} $$
are proportional to the family symmetry breaking vacuum values of the SO(10) singlet fields
$\phi^a, \; S^{ab} = S^{ba}, \; A^{ab} = - A^{ba}$ and the $SO_{10}$
adjoint field $45$.

\subsection{ Features of the Model }

\begin{enumerate}
\item Family symmetry breaking above the GUT scale generates the hierarchy of 
fermion masses.   In the symmetric limit, only the third generation can have mass.  The mass hierarchy for the second and first generations is determined by the family symmetry breaking ratios ($\epsilon, \; \epsilon^\prime$).
$$SU_2 \times U_1 \;\;\; \longrightarrow \;\;\; U_1 \;\;\; \longrightarrow \;\;\; {\rm nothing} $$  
\hspace{0.9in}  $\epsilon$  \hspace{0.6in}  $\epsilon^\prime$

   3rd family  $>>$  2nd family  $>>$  1st  family 
\item Some well known patterns for fermion masses are satisfied approximately in the model.  For example, the  Georgi - Jarlskog relation
$ m_s \sim  \frac{1}{3} m_\mu \;\;\; m_d \sim  3 m_e  \;\;\; @ M_G $ results from the vev of the adjoint $ \langle 45 \rangle = (B - L) M_G $ in the 22 matrix element, where 
$B - L$ is baryon number minus lepton number.   The same adjoint is responsible for Higgs doublet-triplet splitting.

\item  Yukawa coupling unification for the third generation is well respected by the data.  In particular, the relation $\lambda_t = \lambda_b = \lambda_\tau = \lambda_{\nu_\tau}$ $= A$ $@ M_G$ gives  the result   $m_t \sim  170 \pm 20$ GeV.
\item  Gauge coupling unification is well respected by the data.
\item  The $SU_2$ family symmetry naturally suppresses flavor violation such as $\mu \rightarrow e \gamma$.
\item  Finally the model has {\bf only 9 arbitrary Yukawa parameters to fit 13 charged fermion masses and mixing angles}.  Hence there are 4 non-trivial predictions. 
\end{enumerate}

\subsection{ Global $\chi^2$ fit to precision electroweak data\cite{3,4}}

We fit the data using a global $\chi^2$ analysis discussed originally in \cite{2}.  This analysis includes 2 loop RG running from  $M_G \rightarrow M_Z$
with the following boundary conditions at $M_G$.
\begin{itemize}  
\item  universal squark and slepton masses --  $m_0$,
\item  universal gaugino masses - $M_{1/2}$,
\item non-universal  $H_u,  \;\; H_d$ masses,
\item  $\epsilon_3 (\equiv \frac{\alpha_3 - \alpha_G}{\alpha_G})$ parameterizes the effect of one loop threshold corrections to the unification of the three gauge coupling constants at $M_G$.  Note the GUT scale is defined as the scale where $\alpha_1 = \alpha_2$.
\end{itemize}
We also include one loop threshold corrections at $M_Z$
to the $Z$ and $W$ masses and the leading corrections to the down-type quark and charged lepton masses proportional to $\tan\beta$.  We use 3 loop QCD + 1 loop QED RG running below $M_Z$.   We demand a consistent electroweak symmetry breaking solution using the one loop improved Higgs potential, including $m_t^4$ corrections.  Finally we construct a {\bf $\chi^2$ function including 20 low energy observables.}   The result of this fit is given in the following table (for details, see\cite{3,4}).
Note, the observables in bold script have the largest pulls.

The parameters $m_0,\;\; M_{1/2}$ and $\mu$ were fixed. All others were varied using Minuit.  The following are the parameters determined at $M_G$ for the $\chi^2$ fit in the table.  

\noindent
(1/$\alpha_G, \, M_G, \, \epsilon_3$) = ($24.52, \, 3.03 \cdot 10^{16} \;\;
{\rm GeV},$ \\ $-4.06$\%), \\
($\lambda, \,$r$, \, \sigma, \, \epsilon, \, \rho, \, \epsilon^\prime$) =
($ 0.79, \, 12.4, \, 0.84, \, 0.011,$ \\ $0.043,\,  0.0031$),\\
($\Phi_\sigma, \, \Phi_\epsilon, \, \Phi_\rho$) =  ($0.73, \, -1.21, \,
3.72$)rad, \\
($m_0, \, M_{1/2}, \, A_0, \, \mu(M_Z)$) = ($1000,\, 300, \, -1431,$ \\
$110$) GeV,\\
($(m_{H_d}/m_0)^2, \, (m_{H_u}/m_0)^2, \, $tan$\beta$) = ($2.23,\, 1.66,$ \\ $53.7$)

\begin{table}[htb]
\caption{Fit to charged fermion masses and mixing angles.} 
$\begin{array}{|l|c|l|}
\hline
{\rm Observable}  &{\rm Data}(\sigma) & Theory  \\
\mbox{ }   & {\rm (masses} & {\rm in\  \ GeV) }  \\
\hline
\;\;\;M_Z            &  91.187 \ (0.091)  &  91.17          \\
\;\;\;M_W             &  80.388 \ (0.080)    &  80.39       \\
\;\;\;G_{\mu}\cdot 10^5   &  1.1664 \ (0.0012) &  1.166     \\
\;\;\;\alpha_{EM}^{-1} &  137.04 \ (0.14)  &  137.0         \\
\;\;\;{\bf \alpha_s(M_Z)}    &  \bf{0.1190} \ (0.003)   &  \bf{0.1174}       \\
\;\;\;\rho_{new}\cdot 10^3  & -0.20 \ (1.1) & +0.314   \\
\hline
\;\;\;M_t              &  173.8 \ (5.0)   &  174.9 \\
\;\;\;m_b(M_b)          &    4.260 \ (0.11)  &    4.331                  \\
\;\;\;M_b - M_c        &    3.400 \ (0.2)   &    3.426                 \\
\;\;\;\bf{m_s}              &  \bf{0.180} \ (0.050)   &  \bf{0.147}          \\
\;\;\;m_d/m_s          &  0.050 \ (0.015)   &  0.0589        \\
\;\;\;Q^{-2}           &  0.00203 \ (0.00020)  &  0.00201                \\
\;\;\;M_{\tau}         &  1.777 \ (0.0018)   &  1.777         \\
\;\;\;M_{\mu}          & 0.10566 \ (0.00011)   & .1057           \\
\;\;\;M_e \cdot 10^3      &  0.5110 \ (0.00051) &  0.5110  \\
 \;\;\;V_{us}         &  0.2205 \ (0.0026)      &  0.2205        \\
\;\;\;V_{cb}         & 0.03920 \ (0.0030)      &  0.0403           \\
\;\;\;\bf{V_{ub}/V_{cb}}    &  \bf{0.0800} \ (0.02)    &  \bf{0.0691}                 \\
\;\;\;\hat B_K          &  0.860 \ (0.08)    &  0.870           \\
\hline
{B(b\!\rightarrow\! s \gamma)\!\cdot\!10^{4}}  &  3.000 \ (0.47) &  2.992  \\
\hline
  \multicolumn{2}{|l}{{\rm TOTAL}\;\;\;\; \chi^2}  2.26
            & \\
\hline
\end{array}$
\end{table}

Of course once the data is fit, there are other predictions of the theory.  As an example we obtain the CP violating parameters measured in B decays $ sin 2 \alpha  = 0.74 ,\;\;  sin 2 \beta = 0.54 ,\;\; sin \gamma = 0.99$ and the SUSY contribution to the anomalous magnetic moment of the muon $a_\mu^{SUSY} = 40 \times 10^{-10} $.  We also obtain the complete spectrum of squark, slepton, Higgs, and gaugino masses.

\subsection{ Neutrinos in Standard Model }

Let's now consider neutrino masses and mixing.  In the Standard Model neutrinos masses are obtained by introducing three sterile neutrinos $\bar \nu$.
The neutrino masses are then given by the expression $  l {\, U_{e}^\dagger \, \lambda_\nu} \, \bar \nu \, H_u    +  \frac{1}{2} \, \bar \nu^T \, {M} \, \bar \nu $ where the second term is an invariant mass term for the sterile neutrinos.
Note, the charged leptons in the weak doublet $l$ are mass eigenstates and $U_e$ is the unitary matrix used to diagonalize the charged lepton mass matrix.
The mass scale $M$ is not constrained and thus could be taken much larger than $M_Z$.  As a consequence, the heavy sterile neutrinos may be integrated out of the theory, leaving behind the effective 3 x 3 neutrino mass matrix given by
 {$m_\nu^{eff}  =  U_{e}^\dagger \, m_\nu  \, M^{-1} \, m^T_\nu \, U_{e}^* $.

\subsection{ Neutrinos in SO(10)}
A Dirac neutrino mass matrix $m_\nu$ given by $ \nu \ {m_\nu} \ {\bar \nu}$
(with $\nu, \;\; \bar \nu$ in the $16$)
has already been fixed by the fits for charged fermions.

The most economical way of obtaining the see-saw (without introducing higher dimension operators) is by adding three SO(10) singlets $N$ into the superpotential as follows -  $16 \; {\bf \overline{16}} \; N  \; + \; \frac{1}{2} \,  N \; {\bf M_N} \; N$ where the field ${\bf \overline{16}}$ obtains a vev in the right-handed neutrino direction and ${\bf M_N}$ is a mass term.  Both are assumed to be of order $M_G$.  This gives the effective mass terms $N \ {\bf V} \ {\bar \nu}$ + $\frac{1}{2} \, N \; {\bf M_N} \; N$.  We then obtain the 9 x 9 neutrino mass matrix at $M_G$ given by
\begin{eqnarray}
& ( \begin{array}{cccc}\; \nu & \;\; \bar \nu & \;\;  N
\end{array})  &
\nonumber\\  &  \left( {\begin{array}{ccc}  0  & m_\nu & 0  \\
                     m_\nu^T & 0 & V \\
                     0 &  V^T & M_N  \end{array}} \right) & \nonumber
\end{eqnarray}
Upon integrating out the massive states $\bar \nu, \;\; N$ we obtain the effective 3 x 3 neutrino mass matrix -
$ m_\nu^{eff}   =  U_e^\dagger \; 
m_\nu \;(V^T)^{-1}\;M_N\; V^{-1}\; m_\nu^T  \;  U_e^*  $.
Now consider the particular SO(10) theory defined earlier.

\subsection{$SO_{10} \times SU_2 \times U_1^n$ and Neutrino Masses\cite{3}}

The three SO(10) singlets are given by $N_a, \;\; N_3$ with $a = 1,2$. The superpotential below is the most general one consistent with the family symmetries.
\begin{eqnarray}
W \supset &  {\overline{16}} \; (N_a \; \chi^a \;\; + \;\; N_3 \; 16_3) 
 + {1 \over 2} \ N_a \; N_b \; {S^{a \; b}} & \nonumber \\
&  + \;\; N_a \; N_3 \; {\phi^a} & \nonumber \end{eqnarray}
We now see that the effective mass matrix ${\bf M_N}$ discussed above is constrained by the family symmetries and has the form
\begin{eqnarray}
  M_N = & \left( \begin{array}{ccc}  0 & 0 & 0 \\
                                0 & S^{22} & \phi^2 \\
                                 0  & \phi^2  &  0 \end{array}\right) &
\nonumber
\end{eqnarray}

We have analyzed several different possibilities\cite{3}.  In the simplest case, we have only three light active neutrinos.  The effective 3 x 3 neutrino mass matrix is given by  
$$ m_\nu^{eff}    =  m' \; U_e^\dagger \; 
\left(\begin{array}{ccc}  0 & 0 & 0 \\
 0 & b & 1  \\
 0 & 1 & 0  \end{array} \right) \; U_e^* $$
There are only two free parameters in this matrix given by the overall scale $m'$ and $b$.  Both may be taken real without affecting the observable neutrino masses and mixing angles.
The zero in the 33 term is due to an accidental cancellation.  However as a result we naturally obtain maximal $\nu_\mu \longrightarrow \nu_\tau$ mixing for any value of $b \sim \frac{\langle S^{22} \rangle}{\langle \phi^2 \rangle}
\le 1$.   Given this neutrino mass matrix, we find that we can fit both {\it atmospheric} and {\it LSND} but {\bf NOT} {\it atmospheric} and {\it solar} neutrino oscillations!!

If we want to fit both atmospheric and solar neutrino oscillations we must add
a {\it light sterile neutrino}.   I'll describe how we do this next, but in the meantime let me just describe the results.  We now find a solution to  {\it atmospheric} neutrino oscillations with maximal $ \nu_\mu \rightarrow \nu_\tau $ 
mixing and a fit to {\it solar} data given by the   {\bf SMA MSW} solution
with $ \nu_e \rightarrow \nu_{sterile}$ oscillations.  Note, that even though we have four neutrinos, we are not able to fit LSND.

\subsection{ Sterile neutrinos in SUSY GUTs }

We now argue that it is natural to have light sterile neutrinos in the MSSM.
 In principle any SUSY GUT could have many SO(10) singlets. The question is only whether they couple to the observable sector.  Consider adding three such singlets $\;\;\; \bar N^a, \;\; \bar N^3$ with the dimensionful couplings
given by
\begin{eqnarray}
W \supset &  {\bf \mu'} \; N_a \; \bar N^a \;\; + \;\; {\bf \mu_3} \;  N_3 \bar N^3 &
\nonumber 
\end{eqnarray}
The mass scales ${\bf \mu'}, \;\; {\bf \mu_3}$ are assumed to be of order the weak scale. Via standard steps we obtain the neutrino mass matrix at $M_G$

\begin{eqnarray}
& ( \begin{array}{cccc}\; \nu & \;\; \bar N & \;\; \bar \nu & \;\;  N
\end{array})  &
\nonumber\\  &  \left( \begin{array}{cccc}  0 & 0 & m_\nu & 0  \\
                     0 & 0 & 0 & {{\tilde \mu}^T} \\
                     m_\nu^T & 0 & 0 & V \\
                     0 & {\tilde \mu} & V^T & M_N  \end{array} \right) & \nonumber \end{eqnarray}
with 
\begin{eqnarray}
\tilde \mu \equiv & \left( \begin{array}{ccc}  \mu' & 0 & 0 \\
                               0 & \mu' & 0 \\
                                 0  &  0  & \mu_3 \end{array}\right). & \nonumber
\end{eqnarray}
The effective light neutrino mass matrix $m_\nu^{eff}$ is given by
\begin{eqnarray}
m_\nu^{eff} = & \tilde U_e^T  \; {\bf m} \; \tilde U_e  &  \nonumber  
\end{eqnarray}  where

\noindent
${\bf m} = $  \hfill 
\begin{eqnarray}
\left( {\begin{array}{cc}
m_\nu\;(V^T)^{-1}\;M_N\; V^{-1}\; m_\nu^T &  - m_\nu \;(V^T)^{-1}\; \tilde \mu\\
  - {\tilde \mu}^T \; V^{-1}\; m_\nu^T & 0  \end{array}}\right) & & \nonumber  
\end{eqnarray}
and
\begin{eqnarray} \tilde U_e = & \left(\begin{array}{cc} U_e & 0 \\
                                               0 & 1 \end{array}\right). &  \nonumber
\end{eqnarray}

Note we now have light sterile neutrinos and all we need do is assume that the scales ${\bf \mu'}, \;\; {\bf \mu_3}$ are of order $M_Z$.  
  RECALL in the MSSM we have another example of a mass term in the superpotential, $\mu$, defined via $\;\;\; W =  {\mu}  \; H_u \; H_d$ which must also be of order $M_Z$.  What ever symmetries are needed to solve the $\mu$ problem in the MSSM could in principle also explain why our new "mu" parameters are also of order $M_Z$.

\subsection{How robust is this neutrino solution?}

The four neutrino solution found above cannot fit LSND data.  It is important to investigate whether this is a true prediction of the theory.  In order to study this we now assume the possibility of {\it non-minimal} family symmetry breaking vevs $ \langle S^{11} \rangle =  {\kappa_1} \langle S^{22} \rangle, \;\;\;  \langle S^{12} \rangle =  {\kappa_2} \langle S^{22} \rangle$ with  the parameters ${\kappa_1, \kappa_2 << 1 }$ \cite{4}.  In the model presented here, we are not able to find a dynamical argument why $\kappa_1 = \kappa_2 \equiv 0$ should be an exact relation.  Moreover  this extension of the original model has little effect on charged fermion fits for sufficiently small values of $\kappa_1,\;\; \kappa_2$.  However we now find several {\it new solutions} 
given below.  In each solution the atmospheric oscillation data is described by
maximal $\nu_\mu \rightarrow \nu_\tau$  mixing.

We have a $3 \; \nu$ solution with three possibilities for solar neutrinos  --  either  SMA MSW; LMA MSW, or Vacuum $ \nu_e \rightarrow \nu_{active}$ oscillations.
In addition we have $4 \; \nu$ and  $5 \; \nu$ solutions with solar data given by SMA MSW   $\nu_e \rightarrow \nu_{sterile}$ oscillations and LSND given by $\nu_e \rightarrow \nu_\mu$.

\section{ Conclusions }

The theoretical framework presented here provides a {compact and precise description of low energy data.  With regards to neutrino masses and mixing
we have two possibilities --
\begin{itemize}
\item  In a ``predictive'' theory (with sufficient symmetry to enforce a small
number of arbitrary parameters) neutrino masses are very constrained.  In this
case, charged fermion masses and mixing angles can strongly constrain the neutrino sector.

For example, an {$SO_{10} \times D_3 \times Z_M$} model
provides a natural framework
for the case of minimal family symmetry breaking vevs \cite{5}.

\item If there are however many free parameters, then  neutrino masses and mixing are poorly constrained by the charged fermion sector.
\end{itemize}

\noindent
Finally we have not presented a complete theory here.  There are still many
unresolved problems :  SUSY breaking, GUT breaking,  family symmetry breaking,  $\mu$ problem,  strong CP problem.

\end{document}